\documentclass[twocolumn,showpacs,preprintnumbers,amsmath,amssymb,superscriptaddress]{revtex4-2}
\usepackage{epsf}
\usepackage{graphicx}
\usepackage{sidecap}
 \usepackage{soul}
 \usepackage{array}
 \usepackage{amsmath}
 \usepackage{amssymb}
 \usepackage{dcolumn}
 \usepackage{epstopdf}
 \usepackage{bm}
 \usepackage{amsmath}
 
\usepackage{gensymb}
\usepackage{color}
\usepackage{hyperref}
\usepackage{soul}
\sethlcolor{green}
\usepackage{bbding}
\hypersetup{
    colorlinks=true,
    citecolor=red,
    linkcolor=red,
    filecolor=blue,   
    urlcolor=blue,
}
 
\def \beq {\begin{equation}}
\def \eeq {\end{equation}}
\renewcommand{\figurename}{\textbf{Fig.}}
\renewcommand{\thefigure}{{\textbf{\arabic{figure}}}}

\def\bibsection{\refname}
\renewcommand{\refname}{\noindent\textbf{References}}

\begin{document}

\title{{Complex electronic structure evolution of NdSb across the magnetic transition}}

\author{Anup Pradhan Sakhya} \thanks{These authors contributed equally in this work.} \affiliation {Department of Physics, University of Central Florida, Orlando, Florida 32816, USA} 
\author{Baokai Wang} \thanks{These authors contributed equally in this work.} \affiliation {Department of Physics, Northeastern University, Boston, Massachusetts 02115, USA}
\author{Firoza~Kabir} \thanks{These authors contributed equally in this work.} \affiliation {Department of Physics, University of Central Florida, Orlando, Florida 32816, USA} 
\author{Cheng-Yi Huang} \affiliation {Department of Physics, Northeastern University, Boston, Massachusetts 02115, USA}
\author{M.~Mofazzel~Hosen}\affiliation {Department of Physics, University of Central Florida, Orlando, Florida 32816, USA}
\author{Bahadur~Singh}\affiliation{Department of Condensed Matter Physics and Materials Science, Tata Institute of Fundamental Research, Mumbai 400005, India}
\author{Sabin~Regmi}\affiliation {Department of Physics, University of Central Florida, Orlando, Florida 32816, USA}
\author{Gyanendra~Dhakal}\affiliation {Department of Physics, University of Central Florida, Orlando, Florida 32816, USA}
\author{Klauss~Dimitri}\affiliation {Department of Physics, University of Central Florida, Orlando, Florida 32816, USA}
\author{Milo~Sprague}\affiliation {Department of Physics, University of Central Florida, Orlando, Florida 32816, USA}
\author{Robert Smith}\affiliation {Department of Physics, University of Central Florida, Orlando, Florida 32816, USA}
\author{Eric~D.~Bauer}\affiliation{ MPA-Q, Los Alamos National Laboratory, Los Alamos, New Mexico 87545, USA}
\author{Filip~Ronning}\affiliation{Institute for Materials Science, Los Alamos National Laboratory, Los Alamos, New Mexico 87545, USA}
\author{Arun~Bansil} \affiliation {Department of Physics, Northeastern University, Boston, Massachusetts 02115, USA}
\author{Madhab~Neupane} \thanks{corresponding author: Madhab.Neupane@ucf.edu} \affiliation {Department of Physics, University of Central Florida, Orlando, Florida 32816, USA}
\date{\today}
 
\begin{abstract}
 
{\indent The rare-earth monopnictide (REM) family, which hosts magnetic ground states with extreme magnetoresistance, has established itself as a fruitful playground for the discovery of interesting topological phases. Here, by using high-resolution angle-resolved photoemission spectroscopy complemented by first-principles density-functional-theory based modeling, we examine the evolution of the electronic structure of the candidate REM Dirac semimetal NdSb across the magnetic transition. A complex angel-wing-like band structure near the zone center along with arclike features at the zone center and the zone corner are observed. This dramatic reconstruction of the itinerant bands around the zone center is shown to be driven by a magnetic transition: Specifically, the Nd 5\textit{d} electron band backfolds at the $\overline{\Gamma}$ point and hybridizes with the Sb 5\textit{p} hole bands in the antiferromagnetic phase. Our study indicates that antiferromagnetism plays an intricate role in the electronic structure of the REM family.}\end{abstract}
\maketitle

\begin{center}\textbf{I. INTRODUCTION}
\end{center}
The rare-earth monopnictide (REM) materials family provides interesting possibilities for realizing exotic Dirac states and giant magnetoresistance (MR) \cite{zeng, tafti, cava_1, wu, niu, alidoust, neupaneNdSb, guo, nayak, lou, nummy, feng}. Several members of this family, such as CeSb, LaBi, CeBi  have been theoretically predicted and experimentally shown to host non-trivial topological electronic states \cite{alidoust, wu, guo, niu, nayak, lou, nummy, feng}. Other members, of this family, however, such as LaSb, YSb, LuBi, YBi, and LaAs, are topologically trivial \cite{zengLaSb, yangYSb, yu, he, pavlosiuk, yang, liang}. Although the heavier members of the REM family assume an antiferromagnetic (AFM) or ferromagnetic (FM) ground state, much of the existing work \cite{wu, niu, nayak, lou, nummy, feng, zengLaSb, alidoust, neupaneNdSb, he, pavlosiuk} has focused on non-magnetic phases  with a few exceptions \cite{liang, takayama, Y.W, F.W, X.D, CeBi, CeBi2, smbi, smbiarpes}. Magnetic phases of REMs present a rich environment for exploring magnetic interactions and Dirac physics \cite{CeBi, CeBi2}. For example, the heavy-fermion system CeSb exhibits an exotic magnetic phase diagram as a function of the magnetic field, known as a devil’s staircase \cite{Osakabe, takayama, kondo}. A challenge in heavy-fermion physics is to understand the relationship between magnetism and electronic structure associated with the interplay of conduction and {\it f} electrons. With such a versatile ground-state, the REM family is drawing increasing interest in the condensed matter community as a playground for exploring emergent phenomena.\\
\indent NdSb has been suggested to host a Dirac-like dispersion at the corners of the Brillouin zone (BZ) in its paramagnetic (PM) phase. It undergoes an AFM transition at 15 K under zero magnetic field \cite{neupaneNdSb}. NdSb thus is an interesting platform for investigating the effects of magnetism on the electronic structure and topology in the REM family. Neutron diffraction experiments on NdSb polycrystals in zero field at 2 K show commensurate AFM ordering with Nd magnetic moments directed along the ordering wavevector \cite{neutron}. Resistivity measurements indicate that NdSb possesses an extreme magnetoresistance of ~$\sim$ 10$^4$ \%. NdSb displays a complex H-T phase diagram at high fields, with multiple first-order transitions \cite{filip}. However, the electronic structure of NdSb in the low-temperature regime (below 15 K) has not been reported. The present paper attempts to do so and helps understand the role of antiferromagnetism in the electronic structure of NdSb.\\
\indent Here, we report the evolution of the complex electronic band structure of NdSb using high-resolution angle-resolved photoemission spectroscopy (ARPES) across the magnetic transition. Interestingly, an angel-wing-like feature appears at the zone center due to the backfolding of the Nd 5\textit{d} electron pocket present at the $\overline{X}$ point in going from the PM to the AFM phase, which leads to hybridization between the Nd 5\textit{d} and Sb 5\textit{p} bands. Our study reveals the presence of Dirac-like states in both the PM and the AFM phases at the center and corners of the BZ. We find that arclike feature appears as the material undergoes a PM to AFM phase transition. 

\begin{figure}
	\includegraphics[width=9cm]{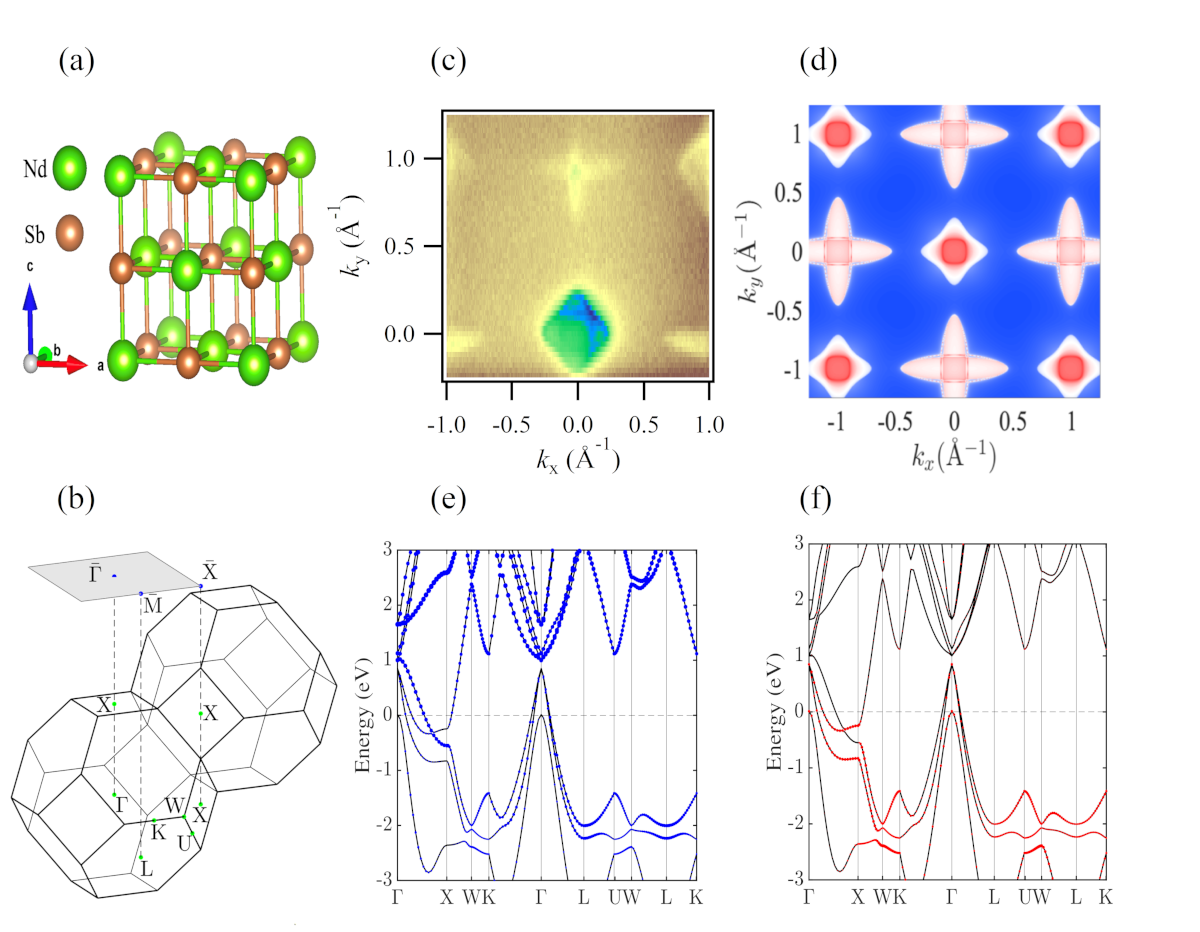} 
	\caption{Crystal and electronic structure of NdSb. (a) Rock-salt type crystal structure of NdSb. (b) Bulk and (001)-surface BZ. (c) Experimental Fermi surface in the PM phase taken at 20 K using a photon energy of 50 eV. (d) Theoretical Fermi surface (with SOC) in the PM phase. (e-f) Band structure (with SOC) along various high-symmetry lines.}
\end{figure}

\begin{center}\textbf{II. EXPERIMENTAL AND COMPUTATIONAL DETAILS}
\end{center}
\indent Single crystals of NdSb were grown by the Sn flux technique as described elsewhere \cite{Growth}. The synchrotron-based ARPES experiments were performed at the ALS beamlines 10.0.1.1 and 4.0.3 equipped with a R4000 and R8000 hemispherical electron analyzer. The angular and energy resolution were set at better than 0.2$^{\circ}$ and 15 meV, respectively. High-quality crystals were cut into small pieces and mounted on a copper post using a Torr seal. Ceramic posts were attached on top of the samples. After loading the sample into the main chamber, the chamber was cooled and pumped down for a few hours. The crystals were cleaved at 20 K and the measurements were carried out over the temperature range of 7-30 K. The pressure in the UHV chamber was better than 1$\times$10$^{-10}$ torr. Electronic structure calculations were performed within the framework of the first-principles, density functional theory (DFT) using the projector augmented wave (PAW) method \cite{PAW} as implemented in the VASP suite of codes \cite{dft, vasp}. The exchange-correlation functional was treated using the strongly constrained and appropriately normed (SCAN) meta-generalized gradient approximation (meta-GGA) \cite{SCAN} functional. An energy cutoff of 400 eV was used for the plane-wave basis set and a $\Gamma$-centered 11$\times$11$\times$11 \textit{k}mesh was used for BZ integrations. The surface energy spectrum was obtained within the iterative Green's function method using the WANNIERTOOLS package \cite{DF, Green2, Green1}. For the antiferromagnetic (AFM) phase, an on-site Coulomb potential U = 14 eV was added to the $\textit{f}$ orbital of Nd to move it away from the Fermi level. We constructed tight-binding models with atom-centered Wannier functions using the VASP2WANNIER90 interface both for the PM and the AFM phases \cite{Marzari}.\\

\begin{center}\textbf{III. RESULTS AND DISCUSSION}
\end{center}
NdSb crystallizes in a rock-salt type crystal structure with space group Fm$\overline{3}$m (Fig. 1(a)) with lattice constant, a = 6.319 \AA~\cite{ndsba}. The bulk BZ of NdSb and its projection on the (001) surface is shown in Fig. 1(b). Figure 1(c) presents the Fermi-surface (FS) map taken at 20 K, which shows the presence of spherical and diamond-shaped pockets around the center $\overline{\Gamma}$ along with two concentric elliptical-electron-like pockets around the corner $\overline{X}$ of the BZ, consistent with the results of previous study \cite{neupaneNdSb}. The theoretical Fermi surface with spin-orbit coupling (SOC) projected on the \textit{k$_x$-k$_y$} plane, is shown in Fig. 1(d), and it is seen to be in excellent agreement with the corresponding experimental results of Fig. 1(c). This agreement implies significant \textit{k$_z$} dispersion of the bulk bands. The associated broadening of the bulk features allows surface states to stand out more clearly in our spectra.

\begin{figure*}
	\centering
	\includegraphics[width=16cm]{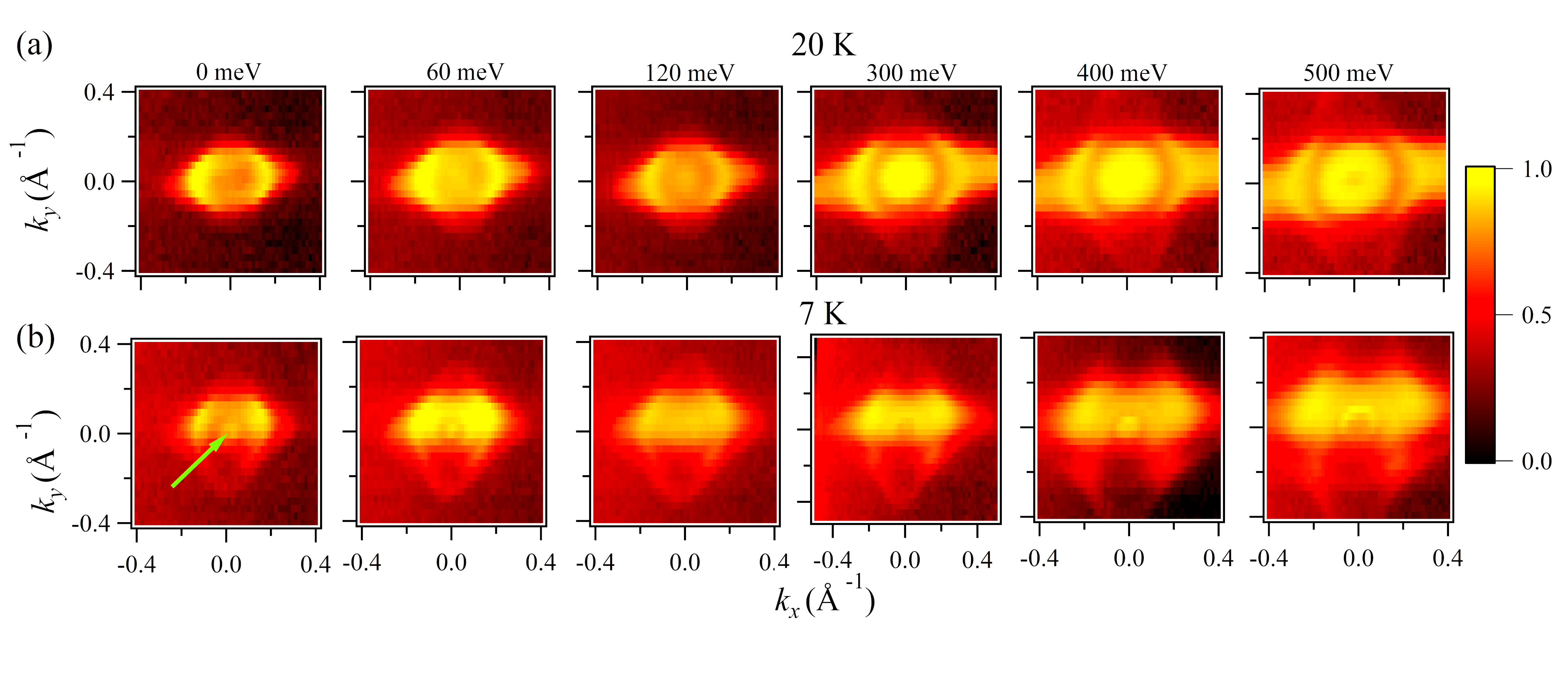}
\vspace{-7ex}
	\caption{Experimental Fermi surface and constant-energy-contours for (a) the paramagnetic phase (20 K) and (b) the antiferromagnetic phase (7 K). Green arrow points to the additional features that appear in the AFM phase. The measurements were performed using a photon energy of 57 eV.}
\end{figure*}

\indent The band structure along the various \textit{k}-directions with SOC is presented in Figs. 1(e) and 1(f), where the blue (red) circles represent contributions from Nd 5\textit{d} (Sb 5\textit{p}) orbitals. One can see that the two hole pockets at the $\Gamma$ point arise mainly from the Sb 5\textit{p} orbitals, whereas the electron pocket at the X point involve a mixture of Nd 5\textit{d} and Sb 5\textit{p} orbitals. From the orbital character plots, we can see the presence of a band inversion between the Nd 5\textit{d} and Sb 5\textit{p} orbitals at the X point. The theoretical analyses of Wilson-loop spectra \cite{Wilson1, Wilson2} at \textit{k$_{z}$} = 0 and \textit{k$_{z}$} =$\pi$ are given in the Supplemental Material (SM) \cite{Sup} and show a strong topological index Z$_2$ = 1, suggesting that the material is topologically nontrivial. 

%\noindent    \textbf{B. Fermi surface evolution due to AFM phase transition} \\
\indent Figures 2(a) and 2(b) present the FS maps and the constant-energy-contours (CECs) in the PM (20 K) and AFM phases (7 K), respectively. A small circular pocket [arrow in Fig. 2(b), left panel], which is absent in the PM phase, appears at the center of the BZ in the AFM phase. This circular feature changes to an almost point like feature at around 60 meV below the Fermi level asserting the electronlike nature of the band. At this binding energy, one can begin to resolve another larger circular pocket, which decreases in size on going further down to 120 meV, where the first circular pocket completely vanishes. Interestingly, at 300 meV below the Fermi level in the AFM phase, the circular pocket appears to be enlarged in size, indicating the holelike nature of the carriers at this binding energy. Moving further towards higher binding energies, the circular pocket enlarges, but the size of this hole pocket is smaller in the AFM phase compared to the PM phase, indicating that the hole band is pushed downwards in energy in the AFM phase.

%\noindent \textbf{C. Evolution of electronic band structure across the AFM transition}\\

\begin{figure*}
\centering
\includegraphics[width=16cm]{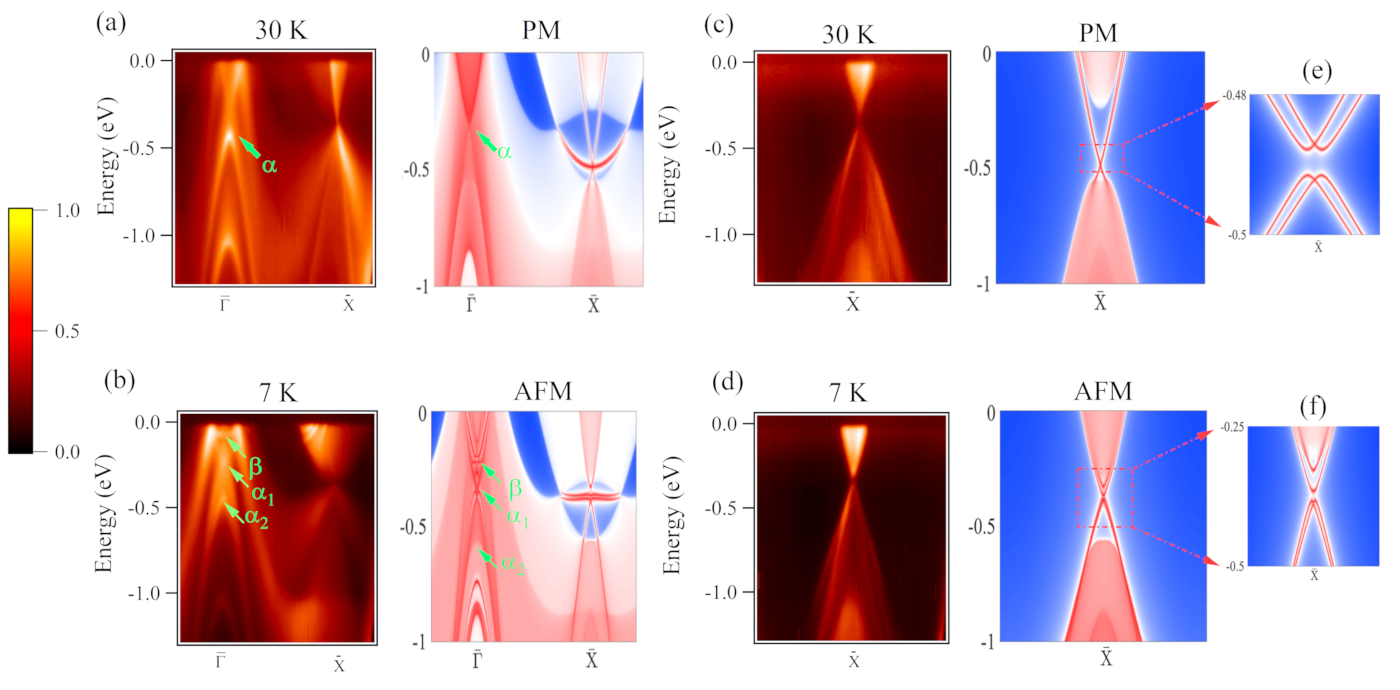}
\vspace{-3ex}
\caption{Evolution of the electronic structure across the AFM transition. Experimental band dispersions along the $\overline{\Gamma}$-$\overline{X}$ direction using ARPES (left panel) and the DFT based (right panel) in (a) the PM and (b) the AFM phase. The green arrows identify the $\alpha$, $\alpha_1$, $\alpha_2$ and $\beta$ bands. Experimental band dispersions along the $\overline{M}-\overline{X}-\overline{M}$ direction (left) and the DFT-based results (right panel) in the (c) PM phase and (d) the AFM phase. The zoomed-in views of the rectangular regions in (c) and (d) are shown in (e) and (f), respectively. The measurements were performed using a photon energy of 50 eV.}
\end{figure*}
\indent In order to explore the effects of magnetism on band dispersions, temperature-dependent measurements along the $\overline{\Gamma}-\overline{X}$ high-symmetry direction were performed for both the PM and the AFM phases (Fig. 3). In the PM phase, there are two holelike bands crossing the Fermi level along with a linear Dirac-like band at the $\overline{\Gamma}$ point marked $\alpha$, (see SM \cite{Sup} for a clearer view). With the onset of antiferromagnetism, significant changes in the electronic structure are observed as shown in Fig. 3(b). In the AFM phase, besides the $\alpha$ (relabeled as $\alpha_1$), we also observed $\alpha_2$ and $\beta$ features. These two features appear inside the hybridization gap resulting from the mixing effect of the folded bands in the AFM phase. The band denoted as $\alpha_1$ shows a signature of a Dirac-like band in the AFM phase. The complex angel-wing-like structure $\beta$ is robust against thermal recycling of the sample, indicating that it reflects effects of the magnetic transition (see SM \cite{Sup} for details). The calculated band structures of the NdSb (001) surface for the PM and AFM phases are compared in Fig. 3 with the ARPES data where the bright red bands are surface states. AFM calculations were performed by taking the ordering wavevector to lie along the $\textit{c}$-axis. A comparison of theory and experiment shows that the observed Dirac-like $\alpha$ bands at the $\overline{\Gamma}$ point in the PM phase, $\alpha_1$ in the AFM phase, and the Dirac-like bands at the $\overline{X}$ point are all surface states (see SM for photon energy dependent measurements in the PM phase \cite{Sup}). Our first-principles calculations reproduce the experimental ARPES data, suggesting that the extra bands at the $\overline{\Gamma}$ point in the AFM phase arise from the zone-folding effect (the detailed band structure with orbital characters given in SM \cite{Sup}, shows that the angel-wing-like structure arises mainly due to the Nd 5\textit{d} states). The Dirac-like bands, [arrows in Fig. 3(b)] have contributions from both the Nd 5\textit{d} and the Sb 5\textit{p} orbitals, suggesting that the Nd 5\textit{d} electronlike pocket at $\overline{X}$ backfolds at $\overline{\Gamma}$ in the AFM phase. We have unfolded the band structure in the AFM phase and compared it with PM bands and find both band structures to display band inversion at the X point of the BZ. No other band inversion was found in the AFM phase, so that the AFM phase is also topologically nontrivial (see SM \cite{Sup}).

\begin{figure}
\includegraphics[width=8cm]{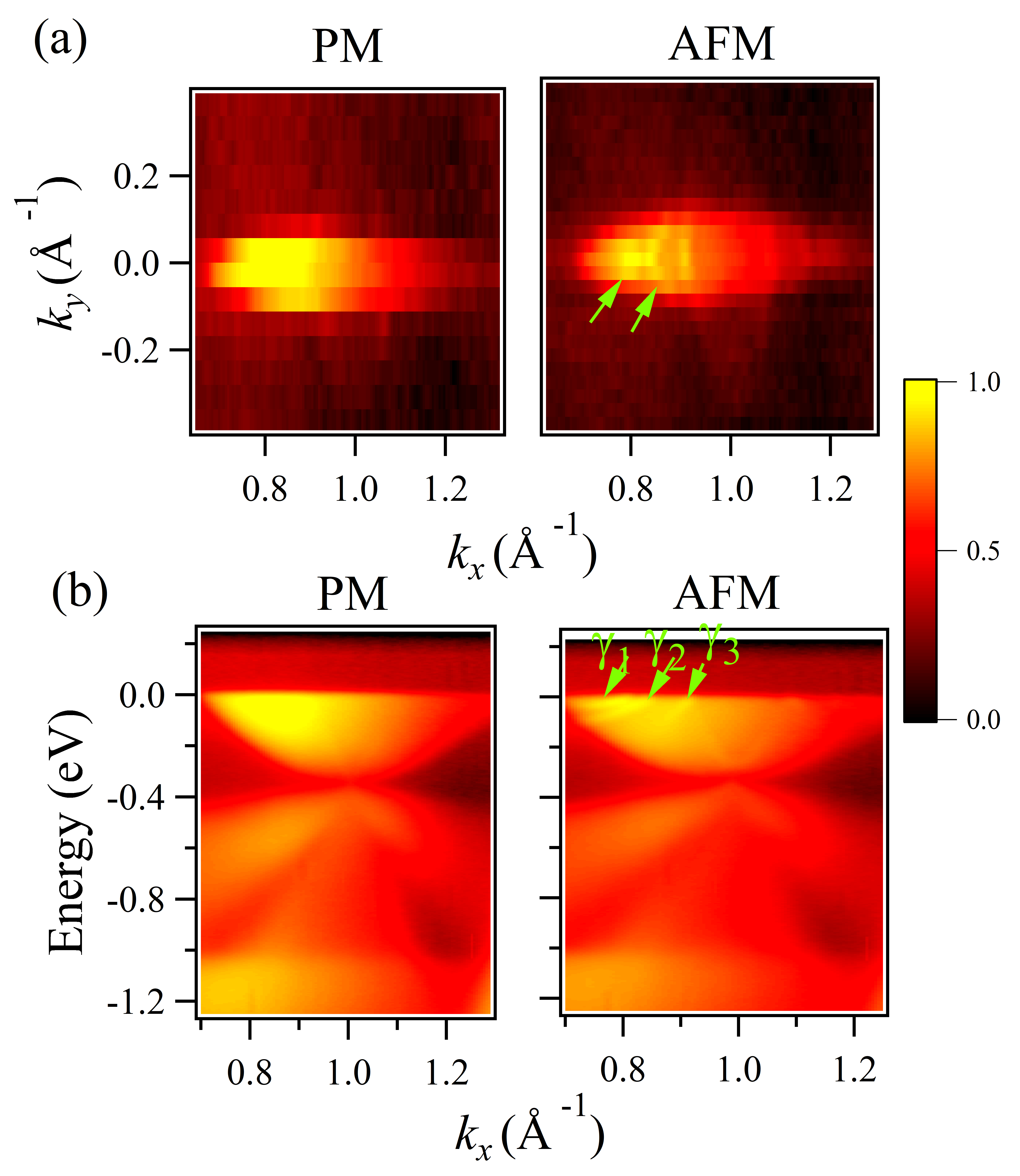}
\vspace{-3ex}
\caption {Effect of the magnetic transition at the corner of the BZ. (a) Fermi surface plots around the zone corner in the PM phase (20 K) and the AFM phase (7 K). (b) Zoomed-in view of the  temperature-dependent dispersion maps around the corner of the BZ in the PM and AFM phases. Dispersion maps were taken along the $\overline{\Gamma}-\overline{X}-\overline{\Gamma}$ direction. Green arrows point to additional bands in the AFM phase. The measurements were performed using a photon energy of 50 eV.}
\end{figure}

\indent From our slab calculations [the right-hand panels of Figs. 3(a) and (3b)], we can see the presence of two surface states at $\overline{X}$ (red-colored bands) within the inverted band gap. The two surface states are tied to the two distinct X points in the bulk BZ when they are projected to the $\overline{X}$ point on the surface BZ. Following Ref \cite{alidoust}, a mirror Chern number of C+ = +1 may be inferred for NdSb since the two surface states at $\overline{X}$ are gapped through hybridization. The presence of one odd surface state at $\overline{\Gamma}$ and two surface states at $\overline{X}$, which is demonstrated in Figs. 3 (a) and 3(b), strongly supports the nontrivial band topology induced by the observed parity inversion. The Dirac-like state at $\overline{X}$ can be seen more clearly when the measurements are carried out along the $\overline{M}-\overline{X}-\overline{M}$ direction as shown in Figs. 3(c,d), in good agreement with a previous report \cite{neupaneNdSb}. Our DFT results, shown in the right-hand panel of Figs. 3 (c) and 3(d), reproduce our ARPES results. Interestingly, our calculations find a small band gap at  $\overline{X}$ [see the zoomed-in view in Figs. 3(e) and 3(f)]. However, the corresponding measured gap at $\overline{X}$ is larger in both the magnetic and the paramagnetic phases. The Fermi level is also slightly shifted in the experiments compared to the DFT calculations. This discrepancy between the experiment and theory likely reflects the inadequacy of the DFT generally in handling low carrier density systems \cite{Gyan, Hosen, Sims}. For gaining more clarity into the effects of the magnetic transition at the corner of the BZ, a zoomed-in view of the Fermi surface at $\overline{X}$ is shown in Fig. 4(a) for the PM and AFM phases, where additional features in the AFM Fermi surface [arrows in Fig. 4(a)] can be seen clearly. An electronlike pocket and a linear Dirac-like state are observed at $\overline{X}$ in the PM as well as the AFM case, and several extra features can be seen near the Fermi level $\gamma_1$,  $\gamma_2$ and $\gamma_3$, which are absent in the PM phase. These features are reproducible with temperature recycling and exist only in the AFM phase (see SM \cite{Sup} for details), but are not produced in our DFT calculations.  Further modeling with a better treatment of the electron correlation effects is needed to get a handle on this discrepancy between theory and experiment at the $\overline{X}$ point. Recently, strikingly similar arclike features have been reported for NdBi, NdSb and CeBi using high-resolution ARPES measurements in the AFM phase \cite{Shrunk, Kushnirenko}. Wang et al. \cite{Wang} were able to reproduce the Fermi arcs in the AFM phase of NdBi using DFT calculations. They discovered that AFM multi-q structures with two (2q) and three (3q) wave vectors gives rise to unconventional surface state pairs on (001) surface of NdBi that is in good agreement with the ARPES data.

\begin{center}\textbf{IV. CONCLUSIONS}
\end{center}
\indent In summary, we have carried out an ARPES study along with parallel first-principles modeling of the electronic and magnetic structures of the rare-earth monopnictide NdSb in the PM and AFM phases. A significant reconstruction of the electronic states is found to take place with the onset of an antiferromagnetic transition. A complex angel-wing-like band structure near the zone center and arclike features at the zone center and the zone corner are observed. The theoretical analysis of Wilson-loop spectra shows a strong topological index Z$_2$ = 1, suggesting that the material is topologically nontrivial. We demonstrate the presence of a linear, Dirac-like band at $\overline{\Gamma}$ and two Dirac-like bands at the $\overline{X}$ point in the BZ in both the PM and the AFM phases, indicating the non-trivial topology of the material. Our study thus highlights the interplay between magnetism and topology in the rare-earth monopnictides.\\
\textit{Note added}: During the reviewing process of our paper, we became aware of a strikingly similar study showing the presence of Fermi arcs in NdBi, NdSb and CeBi in the antiferromagnetic phase \cite{Shrunk, Kushnirenko}.

\begin{center}\textbf{ACKNOWLEDGMENTS}
\end{center}
\indent M.N. acknowledges support from the National Science Foundation (NSF) under CAREER Award No. DMR-1847962, the Center for Thermal Energy Transport under Irradiation, an Energy Frontier Research Center funded by the U.S. DOE, Office of Basic Energy Sciences and the Air Force Office of Scientific Research MURI under Award No. (FA9550-20-1-0322). Work at Los Alamos National Laboratory was carried out under the auspices of the U.S. Department of Energy, Office of Science, Basic Energy Sciences, Materials Sciences and Engineering Division. The work at Northeastern University was supported by the Air Force Office of Scientific Research under award number FA9550-20-1-0322, and  benefited from the computational resources of Northeastern University's Advanced Scientific Computation Center (ASCC) and the Discovery Cluster. This research used resources of the Advanced Light Source, a U.S. Department of Energy Office of Science User Facility, under Contract No. DE-AC02-05CH11231. We thank Sung-Kwan Mo and Jonathan Denlinger for beamline assistance at the Advanced Light Source (ALS), Lawrence Berkeley National Laboratory.\\

\vspace{3ex}
\noindent \textbf{Competing interests}\\
The authors declare no competing interests.\\

\vspace{3ex}
\noindent \textbf{ADDITIONAL INFORMATION}\\

\textbf{Correspondence} and requests for materials should be addressed to Madhab Neupane.
%%%%%%%%%% Merge with supplemental materials %%%%%%%%%%
\clearpage
\widetext
\begin{center}
\textbf{\large Supplemental Materials for \\~\\Complex electronic structure evolution of NdSb across the magnetic transition}
\end{center}
%%%%%%%%%% Merge with supplemental materials %%%%%%%%%%
%%%%%%%%%% Prefix a "S" to all equations, figures, tables and reset the counter %%%%%%%%%%
\setcounter{equation}{0}
\setcounter{figure}{0}
\setcounter{table}{0}
\setcounter{page}{1}
\makeatletter
\renewcommand{\theequation}{S\arabic{equation}}
\renewcommand{\thefigure}{S\arabic{figure}}
\renewcommand{\bibnumfmt}[1]{[#1]}
\renewcommand{\citenumfont}[1]{#1}
\renewcommand{\figurename}{{Supplementary Fig.}}
\renewcommand{\thefigure}{{{\arabic{figure}}}}
\renewcommand{\tablename}{Supplementary Table}
\renewcommand{\thetable}{\arabic{table}}
\def\bibsection{\refname}
\renewcommand{\refname}{\noindent\textbf{Supplementary References}}

\begin{center} \textbf{DFT-BASED BAND STRUCTURES}
\end{center}
\indent The calculated bulk band structures along the high symmetry directions without and with the inclusion of spin-orbit coupling (SOC) are presented in Supplementary Fig. 1. In the absence of SOC, three hole-like bands cross the Fermi level near the $\Gamma$ point as shown in Supplementary Fig. 1(a). An electron-like band can be seen crossing the Fermi level near the X point. Since Nd is a heavy element ($Z$ = 60), SOC is expected to play an important role in the electronic properties. In order to see the effect of SOC on the band structure of NdSb, we have performed calculations by switching on the SOC as shown in Supplementary Fig. 1(b). With the inclusion of SOC, we can observe two important changes in the band structure. Firstly, the inclusion of SOC pushes one of the hole pockets below the Fermi level, and secondly, one can see a band inversion along the $\Gamma$-X direction between the Nd 5\textit{d} and Sb 5\textit{p} orbitals.\\

\begin{figure*} [h!]
\centering
\includegraphics[width=\textwidth]{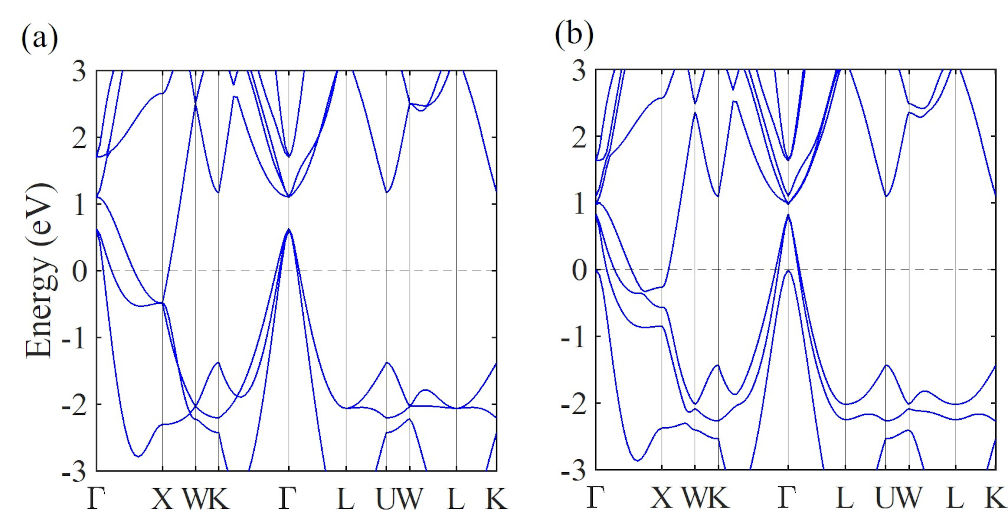}
\caption{The calculated electronic band structure along the high symmetry points (a) without including SOC and (b) with the inclusion of SOC.}
\end{figure*}

\begin{center} \textbf{CALCULATED EIGENVALUES ON THE WILSON LOOPS}
\end{center}
\indent We have checked the topology of NdSb in the PM phase by using the Wilson loop (WL) method, which allows us to trace the topology of a material graphically. Following Ref. \cite{Wilson2r}, we study the evolution of Wannier centers for NdSb in the PM phase. The Wannier centers cross the reference line an odd and even number of times in \textit{k$_{z}$} = 0 and \textit{k$_{z}$} =$\pi$ planes, respectively, as shown in Supplementary Fig. 2, indicating that the system is topologically nontrivial.

\begin{figure*} [h!]
	\centering
	\includegraphics[width=7.3cm]{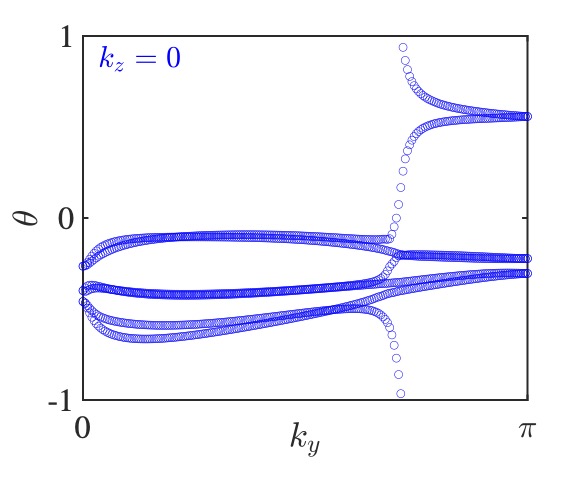}
	\includegraphics[width=7cm]{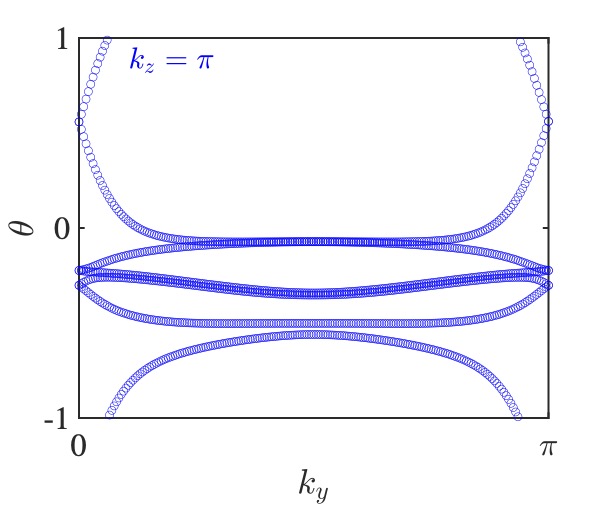}
\caption{The eigenvalues of the Wilson loops along the \textit{k$_{y}$} axis at a fixed \textit{k$_{x}$} in the \textit{k$_{z}$} = 0 and \textit{k$_{z}$} = $\pi$ planes. }
\end{figure*}

\begin{center} \textbf{BAND DISPERSIONS AROUND THE {$\overline{\Gamma}$} POINT}
\end{center}
\indent Supplementary Fig. 3 shows band dispersions around the zone center in the PM and AFM phases. As discussed in the main text, the Dirac-like band is clearly visible in both the PM and the AFM phase as shown in Supplementary Fig. 3, along with the complex angel-wing like structure near the Fermi level (E$_F$). The MDC plots (middle panel) and the second derivative (rightmost panel) of band dispersions of the PM and the AFM phases are also shown for clarity, which further substantiates the features discussed above.\\

\begin{figure*} [h!]
	\centering
	\includegraphics[width=15cm]{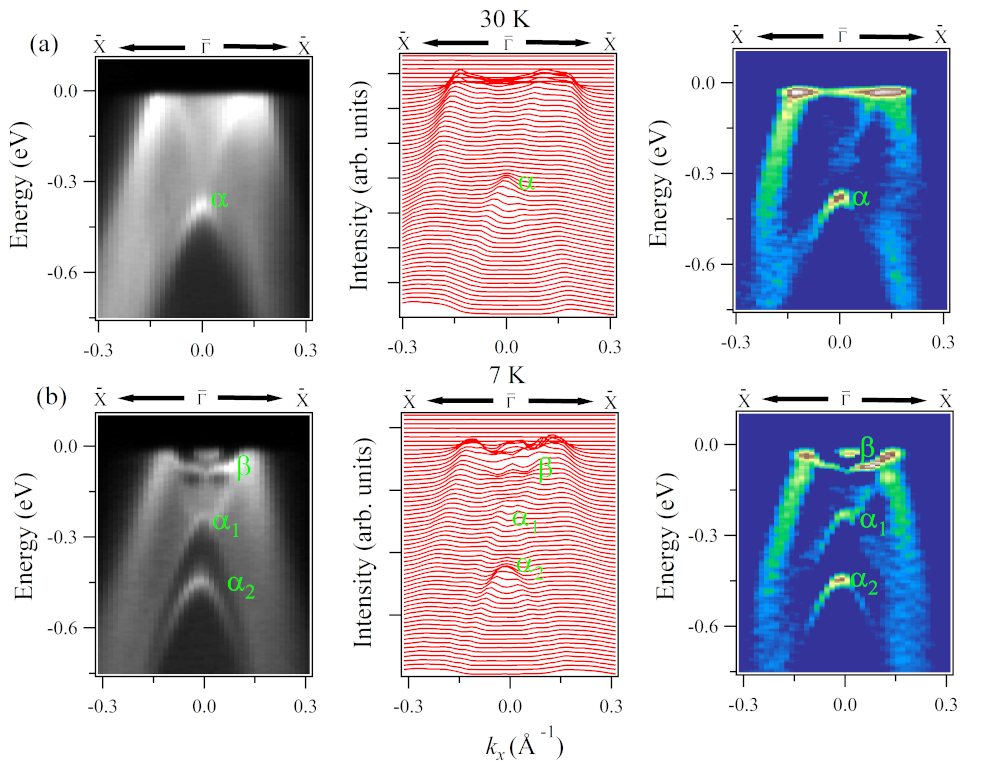}
\caption{Band-evolution around the $\overline{\Gamma}$-point. (a), (b) Experimentally measured band dispersions around the zone center ($\overline{\Gamma}$) (left panel), momentum distribution curves (MDCs) (middle panel), and the second derivative plots using the curvature method (right panel) of paramagnetic (30 K) and antiferromagnetic state (7 K), respectively. The measurements were performed using a photon energy of 57 eV.}
\end{figure*}

\begin{center} \textbf{ELECTRONIC STRUCTURE EVOLUTION ACROSS THE AFM TRANSITION}
\end{center}
\indent The band dispersion obtained from ARPES along the $\overline{\Gamma}$-$\overline{X}$ line is shown in Supplementary Fig. 4. One can clearly see a dramatic reconstruction of the energy bands in the AFM phase compared to the PM phase. This reconstruction of the bands at the $\overline{\Gamma}$ point is due to the backfolding of the Nd 5\textit{d} electron pocket at the $\overline{X}$ point. The three extra features in the AFM phase at the $\overline{X}$ point are clearly seen in Supplementary Fig. 4(b) and cannot be explained by the magnetic zone folding effect. The band reconstruction in the AFM phase at the $\overline{\Gamma}$ and the $\overline{X}$ high symmetry points is found to exist even when the temperature is recycled, suggesting the robustness of the electronic structure of this material.\\

\begin{figure*} [h!]
	\centering
	\includegraphics[width=16cm]{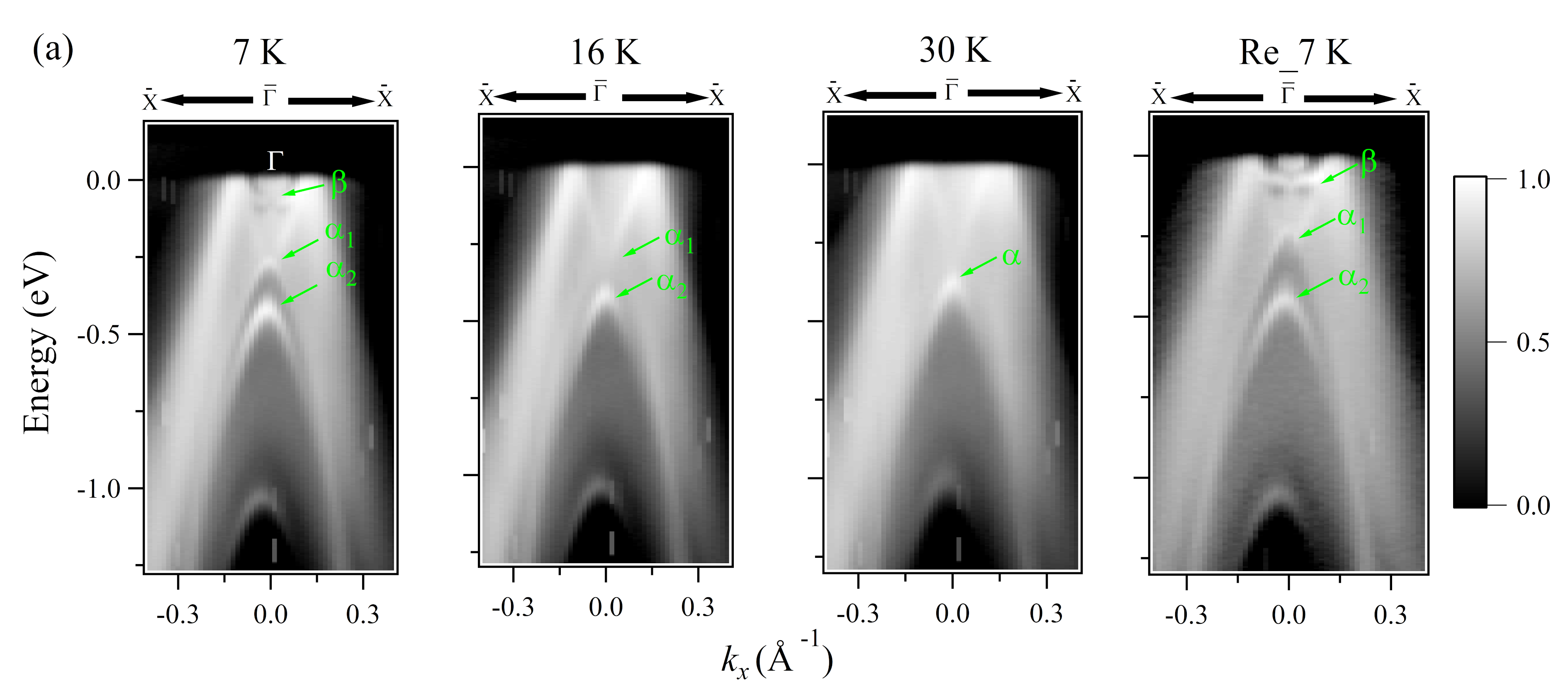}
    \includegraphics[width=15cm]{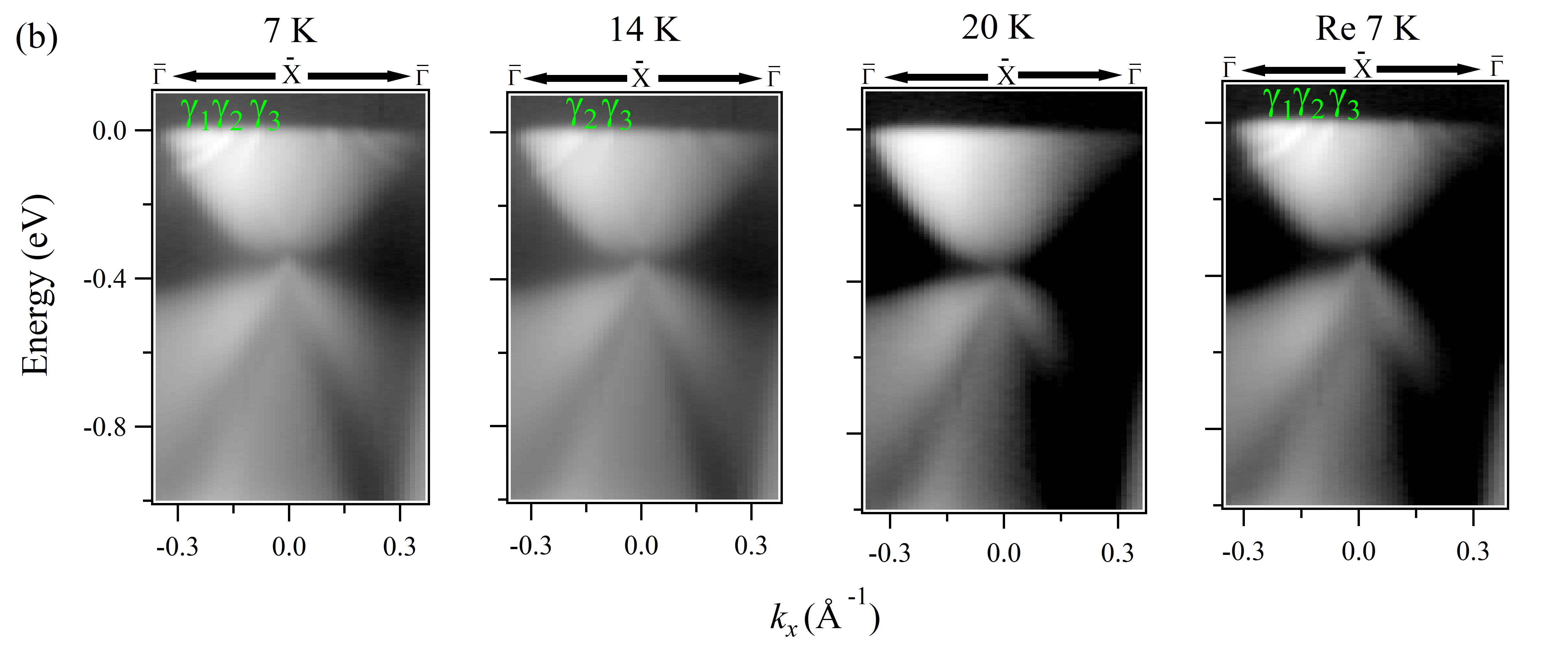}
\caption{ELECTRONIC STRUCTURE EVOLUTION ACROSS THE AFM TRANSITION. (a) Temperature dependent dispersion maps along the $\overline{\Gamma}$-$\overline{X}$ direction. (b) Zoomed in view of the temperature dependent dispersion maps around the corner of the BZ. The dispersion maps were taken along the $\overline{\Gamma}$-$\overline{X}$ line. $\overline{X}$ point is shifted to zero. Temperature values are noted on the plots. The measurements were performed using a photon energy of 57 eV.}
\end{figure*}

\begin{center} \textbf{ORBITAL-PROJECTED SURFACE BAND STRUCTURE IN THE AFM PHASE}
\end{center}
\indent The orbital projected surface band structure of Nd 5\textit{d} and Sb 5\textit{p} spin-up and spin-down states along the $\overline{\Gamma}$-$\overline{X}$ line is shown in Supplementary Fig. 5. The orbital character plot shows that the angel-wing-like structure $\beta$ arises mainly from the Nd 5\textit{d} state whereas the bands ${\alpha_1}$ and ${\alpha_2}$ possesses contributions from both the Nd 5\textit{d} and Sb 5\textit{p} states.\\

\clearpage
\begin{center} \textbf{ELECTRONIC STRUCTURE EVOLUTION AROUND THE $\overline{X}$ POINT}
\end{center}
\indent Supplementary Fig. 6 shows a zoomed in view of the electronic structure evolution near the zone corner in the PM and AFM phases. The middle panel of Supplementary Fig. 6(b) shows the momentum distribution curves, which confirms the existence of three additional features  $\gamma_1$,  $\gamma_2$ and  $\gamma_3$ marked by solid black arrows in the dispersion map. The intensity peaks for all three features are located within 150 meV below the Fermi level. The second derivative plots are shown in the rightmost panel. Apart from these three features, the second derivative plots also reveal clearly a Dirac-like feature at around $\sim$300 meV (see the rightmost panel of Supplementary Figs. 6(a) and (b)).\\

\begin{figure*} [h!]
	\centering
	\includegraphics[width=16cm]{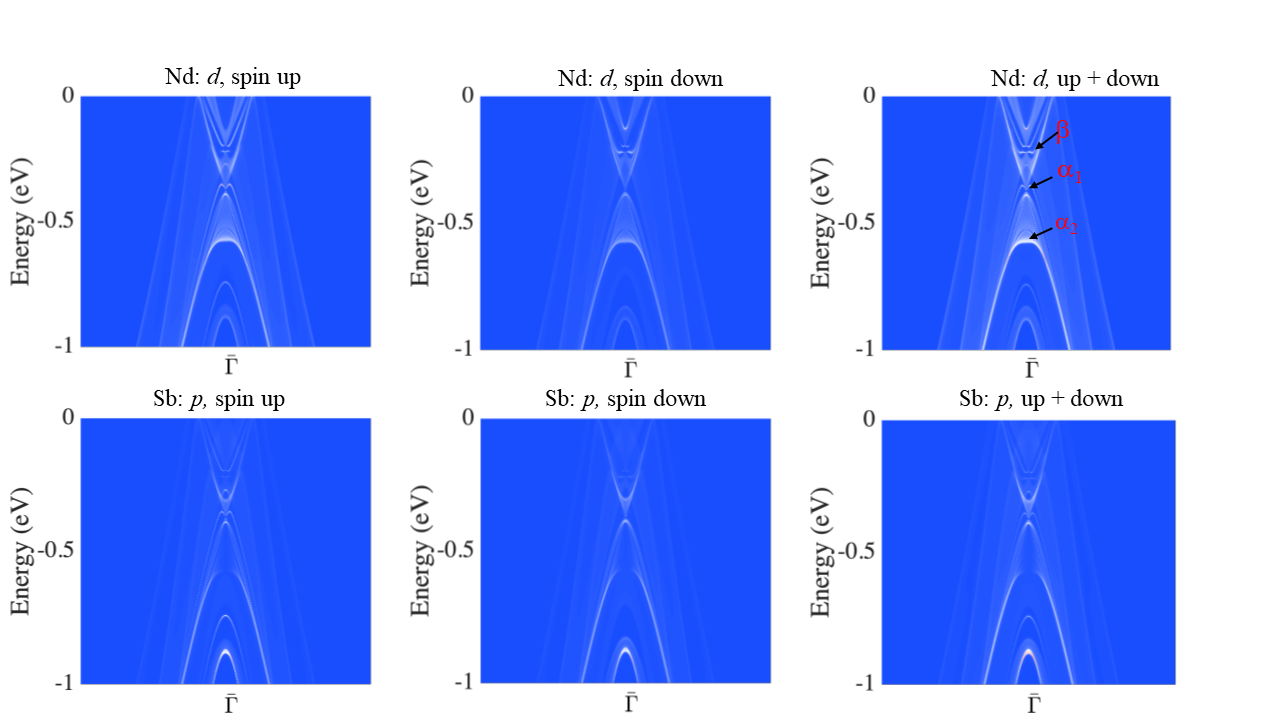}
\caption{Orbital-projected surface band structure of Nd-\textit{d} and Sb-\textit{p} states along the $\overline{\Gamma}-\overline{X}$ high symmetry line n the AFM phase.} 
\end{figure*}

\begin{figure*} [h!]
	\centering
	\includegraphics[width=15cm]{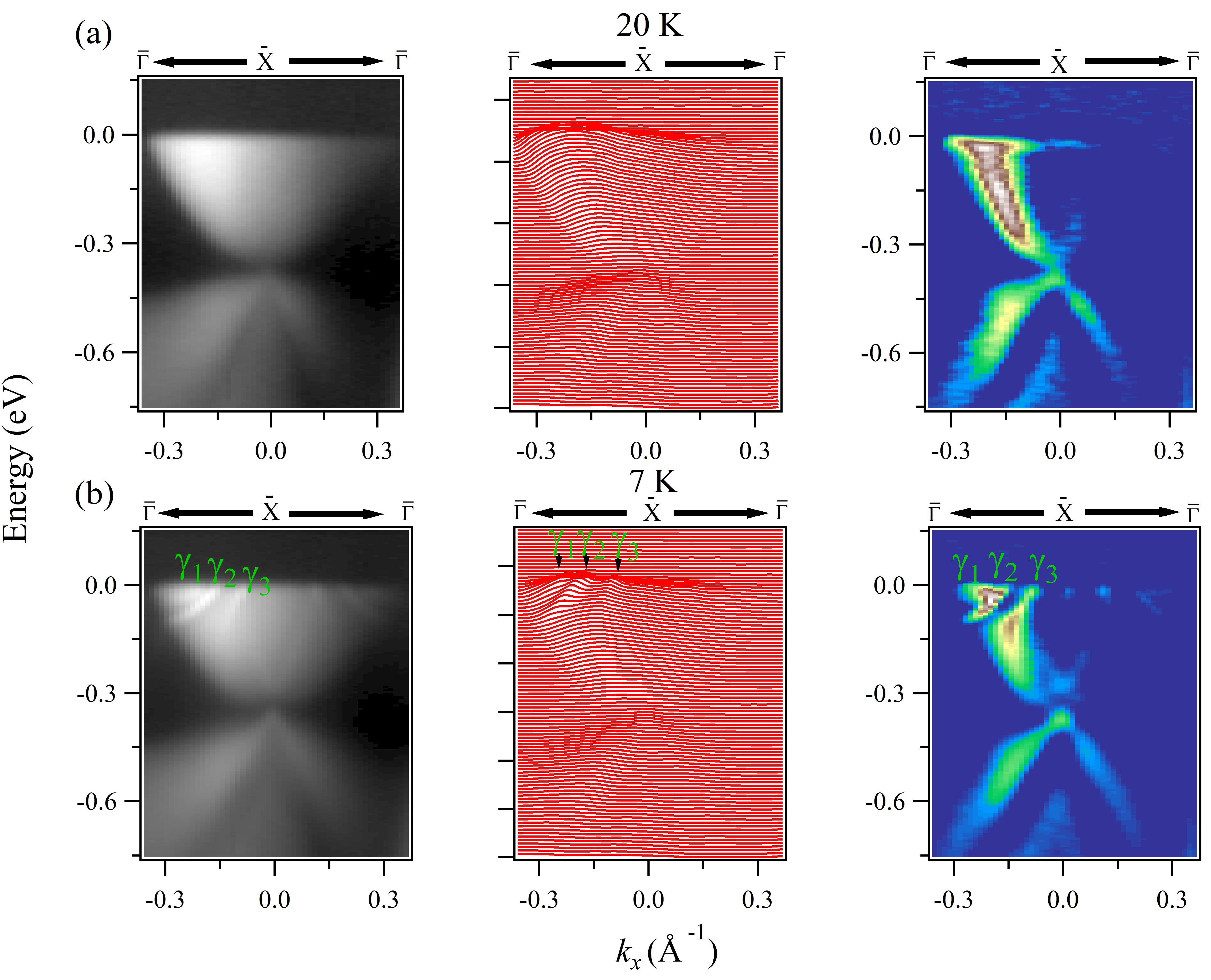}
\caption{Electronic structure around the $\overline{X}$ point. (a),(b) Zoomed-in view of the experimentally measured band dispersion around the zone corner $\overline{X}$ (left panel), MDCs (middle panel) with black arrows marking additional features in the AFM phase, and second derivative plots using the curvature method (right panel) for the  paramagnetic (20 K) and antiferromagnetic state (7 K), respectively. The $\overline{X}$  point is shifted to zero. The measurements were performed using a photon energy of 57 eV.} 
\end{figure*}

\begin{center} \textbf{UNFOLDED BAND STRUCTURE}
\end{center}

\indent Supplementary Fig. 7 shows the unfolded band structure in the PM and the AFM phase where the band inversion at X can be easily visualized thus suggesting non-trivial topology in this material in both the PM and the AFM phase. 

\begin{figure*} [h!]
	\centering
	\includegraphics[width=15cm]{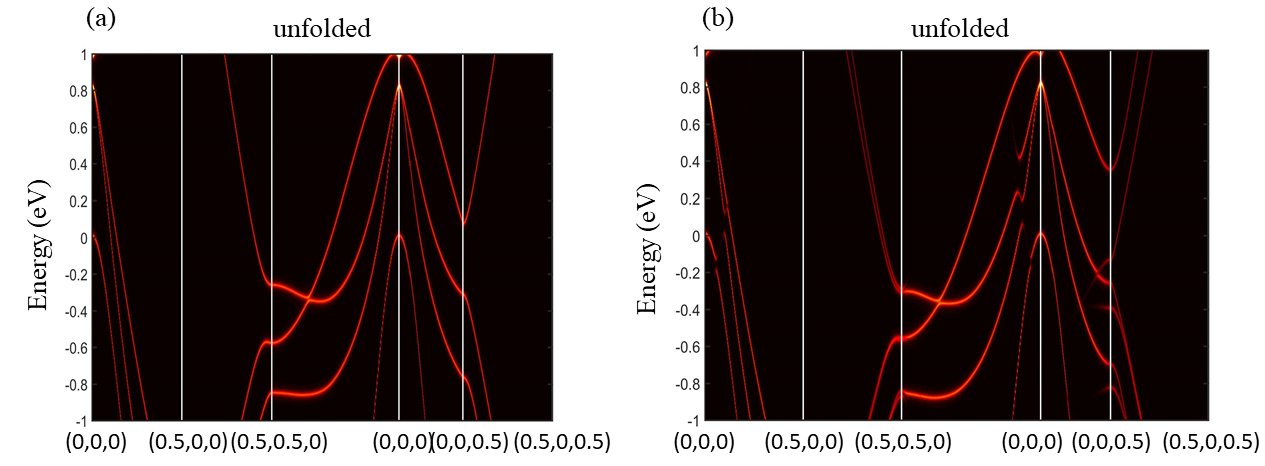}
\caption{Unfolded bandstructure of NdSb in the (a) PM phase and (b) AFM phase.} 
\end{figure*}

\begin{center} \textbf{PRESENCE OF ARCS AROUND THE $\overline{\Gamma}$ AND THE $\overline{X}$ HIGH SYMMETRY POINT}
\end{center}
\noindent Supplementary Fig. 8 shows the electronic structure near the zone corner and the zone center in the AFM phase. The presence of arcs at the $\overline{\Gamma}$ and the $\overline{X}$ high symmetry point can be observed in Supplementary Fig. 8(a) which is more clearly visible in the second derivative plot as shown in Supplementary Fig. 8(b). A narrow window is chosen for further visualization of these arc like features at the $\overline{\Gamma}$ symmetry point as shown in Supplementary Fig. 8(c-d) and at the $\overline{X}$ high symmetry point as shown in Supplementary Fig. 8 (e-f) which reveals the arc like features $\delta_1$, $\delta_2$, $\gamma_1$, and $\gamma_2$ very clearly. These observations are strikingly similar to a recent ARPES report on NdBi, NdSb and CeBi \cite{Shrunkr, Kushnirenkor}. 
\\

\begin{center} \textbf{\textit{k$_{z}$} DEPENDENT MEASUREMENTS}
\end{center}
\indent To conclusively prove the existence of surface states we have presented photon energy dependent measurements along the $\overline{\Gamma}$-$\overline{X}$-$\overline{\Gamma}$ high symmetry direction using photon energy ranging from 30 eV to 120 eV i.e., from {\textit{k$_{z}$} = 3.22 \AA$^{-1}$  to 5.93 \AA $^{-1}$. We have used V$_0$ = 14 eV to calculate the out of the plane momentum. The measurements have been performed at 20 K i.e, in the paramagnetic phase. Supplementary Fig. 9 shows the \textit{k$_{x}$}-\textit{k$_{z}$} maps at different binding energies. The surface states present at the $\overline{\Gamma}$ and the $\overline{X}$ point does not disperse with photon energies suggesting their two-dimensional nature. The photon energies used in the measurements provide extremely surface sensitive measurements. 

\begin{figure*}
\centering
\includegraphics[width=15cm]{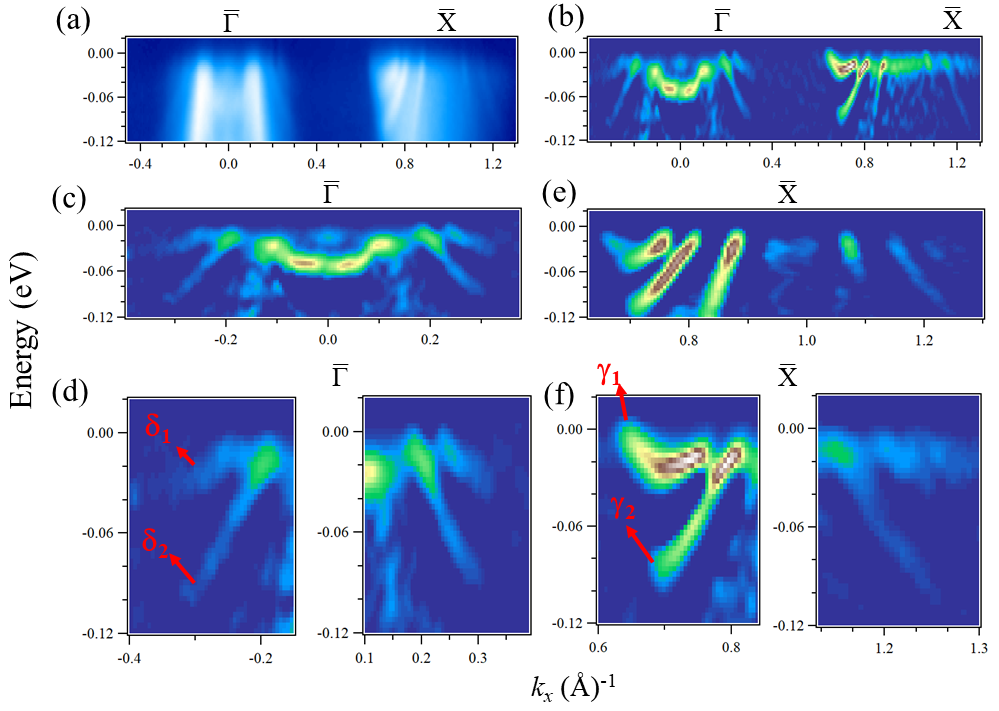}
\caption{(a) Band dispersion along the $\overline{\Gamma}-\overline{X}$ high symmetry direction (b) and its corresponding second derivative plot using curvature method in the AFM phase. Zoomed-in view of the band dispersion around the (c-d) $\overline{\Gamma}$ and (e-f) $\overline{X}$ high symmetry points respectively. The measurements were performed using a photon energy of 50 eV.}
\end{figure*}

\begin{figure*} [h!]
\centering
\includegraphics[width=15cm]{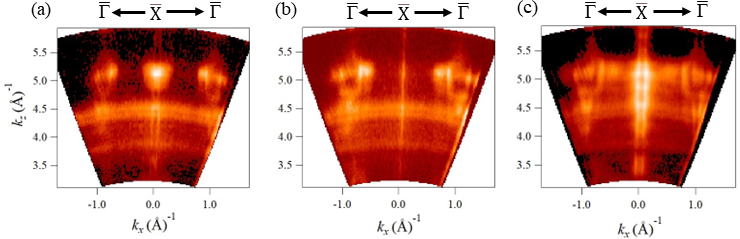}
\caption{\textit{k$_z$} dependent measurement along the $\overline{\Gamma}-\overline{X}$-$\overline{\Gamma}$ high symmetry direction at the binding energy of (a) 0 eV, (b) 320 meV and (c) 540 meV respectively.}
\end{figure*}

\clearpage
\section*{Supplementary References:}
\justify
\bibitem{Wilson2r} R. Yu, X. L. Qi, A. Bernevig, Z. Fang, and X. Dai, \href{https://doi.org/10.1103/physrevb.84.075119} {Phys. Rev. B \textbf{84}, 075119 (2011).}

\bibitem{Shrunkr} B. Schrunk, Y. Kushnirenko, B. Kuthanazhi, J. Ahn, L.-L. Wang, E. O’Leary, K. Lee, A. Eaton, A. Fedorov, R. Lou \textit{et al.,} \href{https://doi.org/10.1038/s41586-022-04412-x} {Nature \textbf{603}, 610-615 (2022).}

\bibitem{Kushnirenkor} Y. Kushnirenko, B. Schrunk, B. Kuthanazhi, L.-L. Wang, J. Ahn, E. O`Leary, A. Eaton, S. L. Bud`ko, R.-J. Slager, P. C. Canfield \textit{et al.,} \href{
https://doi.org/10.1103/PhysRevB.106.115112} {Phys. Rev. B \textbf{106}, 115112 (2022).}

\end{document}